\documentclass[aip,jap,twocolumn,reprint,superscriptaddress,showpacs]{revtex4-1}

\usepackage{graphicx}
\usepackage{amsmath}

\begin{document}

 
\begin{flushright}
YITP-14-48
\end{flushright}

\title{Magnetic ordered structure dependence of magnetic refrigeration efficiency}

\author{Ryo Tamura}
\email[]{TAMURA.Ryo@nims.go.jp}
\affiliation{International Center for Young Scientists, National Institute for Materials Science, 1-2-1, Sengen, Tsukuba, Ibaraki 305-0047, Japan}

\author{Shu Tanaka}
\email[]{shu.tanaka@yukawa.kyoto-u.ac.jp}
\affiliation{Yukawa Institute for Theoretical Physics, Kyoto University,
Kitashirakawa-Oiwakecho, Sakyo-Ku, Kyoto 606-8502, Japan}

\author{Takahisa Ohno}
\email[]{OHNO.Takahisa@nims.go.jp}
\affiliation{Computational Materials Science Unit, National Institute for Materials Science, 1-2-1 Sengen, Tsukuba, Ibaraki 305-0047, Japan}

\author{Hideaki Kitazawa}
\email[]{KITAZAWA.Hideaki@nims.go.jp}
\affiliation{Quantum Beam Unit, National Institute for Materials Science, 1-2-1 Sengen, Tsukuba, Ibaraki 305-0047, Japan}

\date{\today}

\begin{abstract}

We have investigated the relation between magnetic ordered structure and magnetic refrigeration efficiency in the Ising model on a simple cubic lattice using Monte Carlo simulations.
The magnetic entropy behaviors indicate that the protocol, which was first proposed in [Appl. Phys. Lett. {\bf 104}, 052415 (2014).], can produce the maximum isothermal magnetic entropy change and the maximum adiabatic temperature change in antiferromagnets.
Furthermore, the total amount of heat transfer under the proposed protocol reaches a maximum.
The relation between measurable physical quantities and magnetic refrigeration efficiency is also discussed.

\end{abstract}



\maketitle


\section{Introduction} \label{sec:intro}

Cooling phenomena are widely used in everyday applications, such as food storage and medical treatment, and also in technology for next-generation electronics, such as hydrogen-fuel cells\cite{Turner-1999,Steele-2001,Schlapbach-2001,Crabtree-2004} and quantum information processing\cite{Nakamura-1999,Leuenberger-2001,Ladd-2010,Johnson-2011}. 
Thus, high-performance cooling technology has been actively developed for many fields. 
A ubiquitous cooling technology is gas refrigeration, which rely on a compressor cycle of refrigerant gas.
Another method, called magnetic refrigeration, which uses magnetic materials for cooling, has attracted much attention, because the magnetic entropy density is high compared with that of gas refrigeration\cite{Warburg-1881,Debye-1926,Giauque-1927,Giauque-1933,Hashimoto-1981,Hashimoto-1987,Bennett-1992,Bennett-1993,Shaw-1994,Shaw-1995,Pecharsky-1999,Novotny-1999,Pecharsky-2001,Zhang-2001,Tishin-2003,Zhitomirsky-2003,Provenzano-2004,Gschneidner-2005,Campos-2006,Zou-2009,Shen-2009,Sakai-2009,Zvere-2010,Oliveira-2010,Mamiya-2010,Buchelnikov-2010,Sadakuni-2010,Lyubina-2010,Zhu-2011,Perez-2012,Franco-2012,Yonezawa-2013,Mizumaki-2013,Lorusso-2013,Moya-2014,Choudhury-2014,Pospisil-2014}.
The magnetic entropy change is caused by varying control parameters, such as the temperature and magnetic field, which is called magnetocaloric effect (MCE), and it leads to magnetic refrigeration. 
The magnetic entropy is low in magnetic ordered states, such as ferromagnetic and antiferromagnetic states. 
However, the magnetic entropy increases as the temperature increases, because the equilibrium state is disordered. 
In a word, high-efficiency magnetic refrigeration can be achieved by designing low- and high-entropy states.

\begin{figure}[!b]
\begin{center}
\includegraphics[scale=1.0]{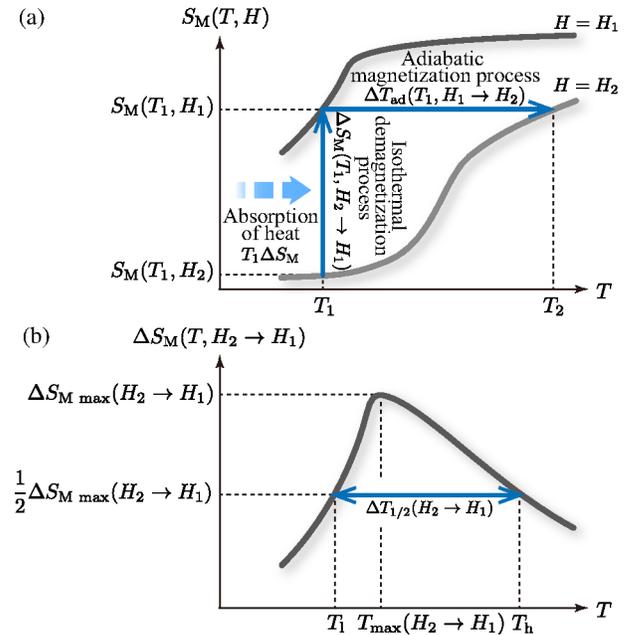} 
\end{center}
\caption{\label{fig:schematic_S}
(Color online)
(a) Temperature dependence of the magnetic entropy $S_\text{M} (T,H)$ under magnetic fields $H_1$ (black line) and $H_2$ (gray line) ($H_1<H_2$) in ferromagnets or paramagnets.
Two typical processes in magnetic refrigeration are indicated by the solid arrows.
(b) Temperature dependence of the isothermal magnetic entropy change $\Delta S_\text{M} (T,H_2 \to H_1)$ in ferromagnets or paramagnets.
}
\end{figure}

The temperature $T$ and magnetic field $H$ dependences of the magnetic entropy $S_\text{M} (T,H)$ are important for designing magnetic refrigeration cycles.
In ferromagnets and paramagnets, the magnetic entropy decreases as the magnetic field increases at a given temperature (Fig.~\ref{fig:schematic_S} (a)).
Figure~\ref{fig:schematic_S} (a) shows the two main processes in magnetic refrigeration.
In the isothermal demagnetization process, when the magnetic field decreases from $H_2$ to $H_1$ at temperature $T_1$, the magnetic entropy changes according to
\begin{align}
\Delta S_\text{M} (T_1,H_2 \to H_1) = S_\text{M} (T_1,H_1) - S_\text{M} (T_1,H_2),
\label{eq:def_DSM}
\end{align}
which is called the isothermal magnetic entropy change.
$\Delta S_\text{M} (T_1,H_2 \to H_1)$ in Fig.~\ref{fig:schematic_S} (a) is positive, which means that the magnetic entropy increases and the magnetic system absorbs the amount of heat $T_1 \Delta S_\text{M} (T_1,H_2 \to H_1)$.
Thus, a large isothermal magnetic entropy change is required for good magnetic refrigeration materials. 
In the adiabatic magnetization process, the magnetic field increases from $H_1$ to $H_2$ with no change in the magnetic entropy. 
The initial temperature is $T_{1}$ and the temperature of the magnetic materials changes according to
\begin{align}
\Delta T_\text{ad} (T_1,H_1 \to H_2) = T_2 - T_1,
\end{align}
where $T_2$ is the temperature, such that $S_\text{M}(T_1,H_1)=S_\text{M}(T_2,H_2)$.
$\Delta T_\text{ad} (T_1,H_1 \to H_2)$ is the adiabatic temperature change.
$\Delta T_\text{ad} (T_1,H_1 \to H_2)$ in Fig.~\ref{fig:schematic_S} (a) is positive, which means that the temperature of the magnetic material increases.
In the active magnetic regenerator (AMR) cycle or its similar cycles\cite{Brown-1976,Patton-1986,Rowe-2006,Zimm-2006,Okamura-2006,Utaki-2007,Fujita-2007,Matsumoto-2011}, 
the adiabatic temperature change is used directly.
Thus, a large adiabatic temperature change is another requirement for good magnetic refrigeration materials.
The relative cooling power (RCP) has been often used as a benchmark for good magnetic refrigeration materials. \cite{Gschneidner-2000}
The RCP is defined as
\begin{align}
&\text{RCP}(H_2 \to H_1) \notag \\
&\ \ \ \ \ = \Delta S_\text{M max} (H_2 \to H_1) \times \Delta T_{1/2} (H_2 \to H_1), \label{eq:RCP}
\end{align}
where $\Delta S_\text{M max} (H_2 \to H_1)$ and $\Delta T_{1/2} (H_2 \to H_1)$ are the maximum value and the full width at half maximum of $\Delta S_\text{M} (T,H_2 \to H_1)$ at given $H_{1}$ and $H_{2}$, respectively (Fig.~\ref{fig:schematic_S} (b)).
Magnetic materials with a large RCP exhibit a large isothermal magnetic entropy change over a wide temperature range.
Hereafter, the argument of the quantities will be sometimes omitted.

Magnetic materials with a large $\Delta S_\text{M}$, $\Delta T_\text{ad}$, and RCP are considered to have a high magnetic refrigeration efficiency.
Ferromagnets are a good magnetic refrigeration material because of their large magnetic field response near the Curie temperature\cite{Pecharsky-1997,Dankov-1998,Wada-2001,Wada-2002,Tegus-2002,Fujieda-2002,Fujita-2003,Fujieda-2006,Yan-2006,Chikazumi-2009,Lyubina-2012}.
In a simple protocol, the magnetic field changes from finite $H$ to zero during the isothermal demagnetization process, and from zero to finite $H$ during the adiabatic magnetization process; that is, $H_1=0$ in Fig.~\ref{fig:schematic_S}.
This protocol has been used in most previous studies of magnetic refrigeration.
We refer to this method as the conventional protocol.
Recently, the MCE in non-ferromagnetic materials, such as antiferromagnets and random magnets, has been experimentally investigated to explore their potential application in magnetic refrigeration\cite{Bohigas-2002,Sosi-2005,Roy-2006,Luo-2006,Luo-2007,Samanta-2007a,Samanta-2007b,Hu-2008,Li-2009a,Li-2009b,Chen-2009,Naik-2011,Kim-2011,Yuan-2012}.
In these magnetic materials, the magnetic refrigeration efficiency calculated with the conventional protocol is sometimes small.

We have focused on the relation between the magnetic refrigeration efficiency and the magnetic ordered structure.
In Ref.~\onlinecite{Tamura-2014}, we focused on just the isothermal demagnetization process and investigated $\Delta S_\text{M}$ of the Ising model on a simple cubic lattice using the Wang-Landau method. \cite{Wang-2001a, Wang-2001b,Lee-2006}
Ref.~\onlinecite{Tamura-2014} studied $\Delta S_\text{M}$ in a ferromagnet and in A-, C-, and G-type antiferromagnets, which are typical magnetic ordered structures.
The MCE in the antiferromagnets differs from that in the ferromagnet.
Furthermore, we proposed a new magnetic refrigeration protocol.
The proposed protocol produces larger $\Delta S_\text{M}$ in antiferromagnets than the conventional protocol.

This paper continues the work in Ref.~\onlinecite{Tamura-2014} and aims to elucidate the relation between the magnetic refrigeration efficiency and typical magnetic ordered structures.
In particular, we focus on the following.
(i) We study the relation between MCE and measurable physical quantities, such as specific heat and magnetization.
(ii) We consider the performance of magnetic refrigeration in both the isothermal demagnetization process and the adiabatic magnetization process.
 
The rest of the paper is organized as follows.
In Sec.~\ref{sec:model}, we introduce the Ising model on a simple cubic lattice and the magnetic ordered structures considered in this paper.
In Sec.~\ref{sec:MC}, we explain how to obtain physical quantities using the Wang-Landau method.
The dependence of the magnetic entropy on the magnetic ordered structures is shown.
In addition, we discuss the relation between the magnetic entropy and the behavior of the specific heat and magnetization.
In Sec.~\ref{sec:MCE}, the magnetic refrigeration efficiency in the isothermal and adiabatic magnetization processes is considered.
To estimate the efficiency, we calculate the isothermal magnetic entropy change and the adiabatic temperature change.
The RCP is not sufficient for estimating the magnetic refrigeration performance in this study.
Thus, to investigate the efficiency in the isothermal demagnetization process in more detail, we introduce a new quantity called total cooling power (TCP) to replace the RCP.
Section \ref{sec:conclusion} is the conclusion.
In Appendix~\ref{sec:TCP}, we explain the properties of the TCP.
Mean-field analysis of magnetic refrigeration is described in Appendix~\ref{sec:MF}.


\begin{figure*}[t]
\begin{center}
\includegraphics[scale=1.0]{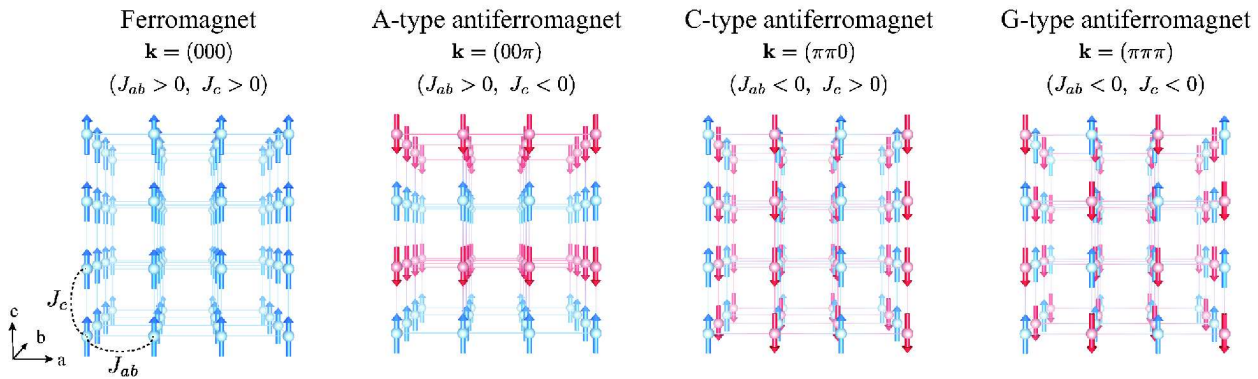} 
\end{center}
\caption{\label{fig:structure}
(Color online)
Schematics of ordered magnetic structures in the Ising models on a simple cubic lattice.
The signs of the magnetic interactions $J_{ab}$ and $J_c$ and the wave vector $\mathbf{k}$, when the lattice constant is set to unity, are shown for each magnetic structure.
These figures were drawn by VESTA\cite{Momma-2011}.
}
\end{figure*}

\section{Ising model on a simple cubic lattice} \label{sec:model}

In this section, we introduce the model for investigating the dependence of the magnetic refrigeration efficiency on the magnetic ordered structures.
We consider the MCE in the $S=1/2$ Ising models on a simple cubic lattice using statistical thermodynamics.
Let $N= L \times L \times L$ be the number of sites, where $L$ is the linear dimension. 
The model Hamiltonian is defined by
\begin{align}
\mathcal{H} = - J_{ab} \sum_{\langle i,j \rangle_{ab}} s_i^z s_j^z 
                      - J_{c} \sum_{\langle i,j \rangle_{c}} s_i^z s_j^z
                      - H \sum_{i} s_i^z,
                      \ \ \ s_i^z = \pm \frac{1}{2}, 
\label{eq:model}
\end{align}
where $J_{ab}$ and $J_{c}$ are nearest-neighbor interactions in the $ab$-plane and in the $c$-axis, respectively, and $H$ is the uniform magnetic field parallel to the $z$-axis of spin.
Here, the $g$-factor and the Bohr magneton $\mu_\text{B}$ are set to unity.
The periodic boundary conditions are imposed for all directions.
In this paper, we focus on the ferromagnetic structure and the A-, C-, and G-type antiferromagnetic structures shown in Fig.~\ref{fig:structure} as in Ref.~\onlinecite{Tamura-2014}.
These are typical magnetic ordered structures.

Throughout this paper, the absolute values of interactions are set as the same, $J := |J_{ab}|=|J_{c}|$, where $J$ is the energy unit.
The A-, C-, and G-type antiferromagnetic structures are bipartite magnetic structures. 
The blue and red arrows in Fig.~\ref{fig:structure} indicate the spins on respective sublattices. 
At $H=0$, the thermodynamic properties of these models, which have any one of the antiferromagnetic ground states, are the same as the ferromagnetic Ising model when the gauge transformation ($s_i^z \to -s_i^z$ for any $i$ in one of sublattices) is applied.
Thus, a second-order phase transition occurs at the critical temperature $T_\text{c}/J=1.127\cdots$\cite{Ferrenberg-1991} for all cases where $H=0$.
Here, the Boltzmann constant $k_\text{B}$ is set to unity.
For the ferromagnet, $T_\text{c}/J$ is the Curie temperature, and $T_\text{c}/J$ is the N\'eel temperature for the antiferromagnets.


\section{Monte Carlo simulation results} \label{sec:MC}


In this section, we consider the MCE behavior in the Ising model. 
We obtain the dependence of the magnetic entropy, the magnetic specific heat, and the magnetization on $T$ and $H$ by using the Wang-Landau method\cite{Wang-2001a, Wang-2001b,Lee-2006}.
The Wang-Landau method is a Monte Carlo method and performs a random walk in energy space.
It can directly calculate the absolute density of states $g(E,H)$, where $E$ is the energy of the state.
The absolute density of states is normalized as $\sum_E g (E,H) = 2^N$ which corresponds to the total number of states.
In other words, the magnetic entropy per spin is $\ln 2=0.693\cdots$ for the limit of $T\to \infty$.
The partition function $Z (T,H)$, the Helmholtz free energy $F (T,H)$, and the internal energy $U (T,H)$ at a given $T$ and $H$ can be calculated with the obtained $g(E,H)$ by
\begin{align}
Z (T,H) &= \sum_{E} g (E,H) \text{e}^{- \beta E}, 
\label{eq:partition}\\
F (T,H) &=- T \ln Z(T,H), \\
U (T,H) &= \frac{1}{Z(T,H)} \sum_{E} E g(E,H) \text{e}^{- \beta E},
\end{align}
where $\beta$ is the inverse temperature $1/T$.
Using these quantities, $S_\text{M} (T,H)$ and the magnetic specific heat per spin $C_\text{M} (T,H)$ are obtained by
\begin{align}
S_\text{M} (T,H) &= \frac{1}{N} \frac{U(T,H)-F(T,H)}{T}, 
\label{eq:entropy} \\
C_\text{M} (T,H) &= \frac{1}{N} \frac{\partial U(T,H)}{\partial T} 
\label{eq:specificheat}.
\end{align}
We can directly calculate the magnetic entropy without integrating the magnetic specific heat or the magnetization, which is an advantage of the Wang-Landau method.
Furthermore, 
the magnetization per spin $m (T,H)$ is calculated by
\begin{align}
m (T,H) = \frac{1}{Z(T,H)} \sum_E \langle \tilde{m} (E,H) \rangle g (E,H) \text{e}^{-\beta E},
\label{eq:mag}
\end{align}
where $\langle \tilde{m} (E,H) \rangle$ is a microcanonical ensemble average of the magnetization per spin, which can be calculated simultaneously with $g (E,H)$.


\begin{figure*}
\begin{center}
\includegraphics[scale=1.0]{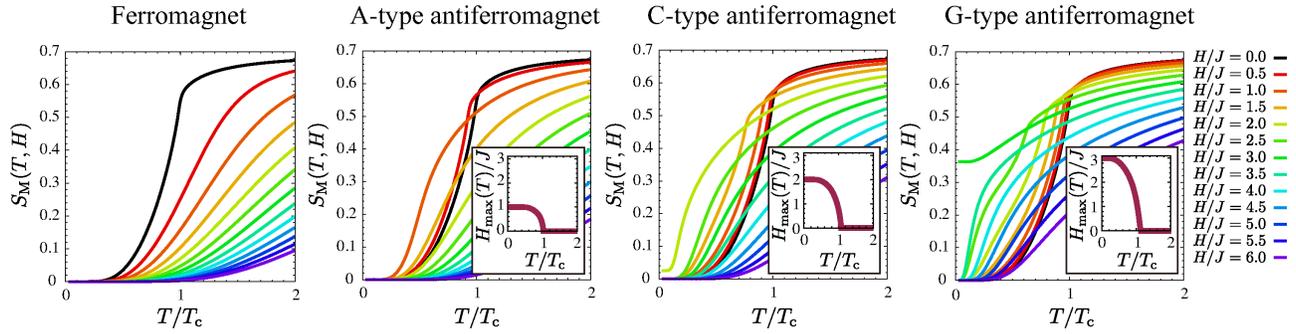} 
\end{center}
\caption{\label{fig:MC_entropy}
(Color online)
Temperature dependence of the magnetic entropy per spin $S_\text{M}(T,H)$ for $L=16$ under various magnetic fields obtained by the Wang-Landau method.
The insets show the temperature dependence of $H_\text{max} (T)$ at which the magnetic entropy reaches its maximum.
}
\end{figure*}
\begin{figure*}
\begin{tabular}{c}
\begin{minipage}{0.3\hsize}
\begin{center}
\vspace{-5mm}
\includegraphics[scale=1.0]{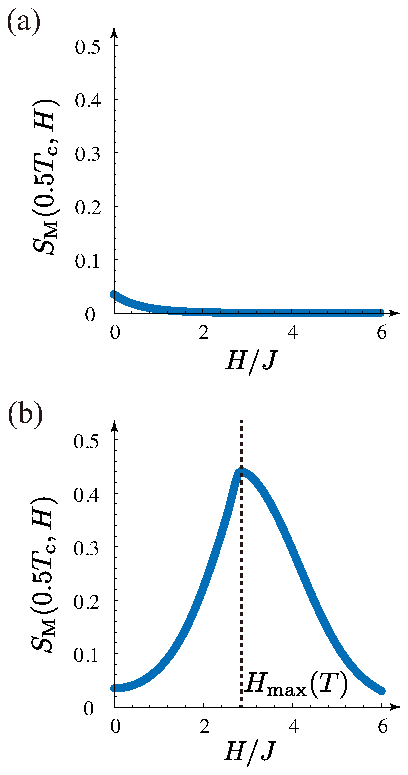} 
\end{center}
\end{minipage}
\begin{minipage}{0.7\hsize}
\begin{center}
\vspace{-5mm}
\includegraphics[scale=1.0]{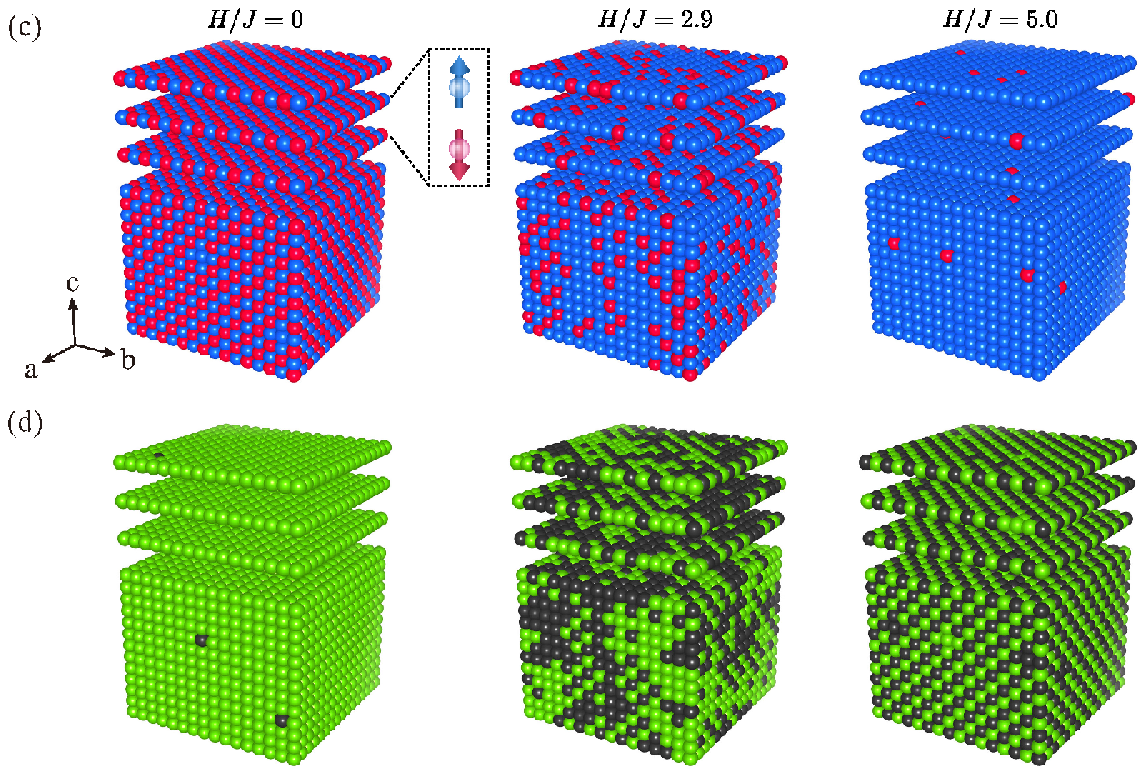} 
\end{center}
\end{minipage}
\end{tabular}
\caption{\label{fig:snapshot}
(Color online)
(a) Magnetic field dependence of $S_\text{M} (T,H)$ with $T/T_\text{c}=0.5$ for $L=16$ for the ferromagnet.
(b) Magnetic field dependence of $S_\text{M} (T,H)$ with $T/T_\text{c}=0.5$ for $L=16$ for the G-type antiferromagnet.
(c) Snapshots of the spins of the G-type antiferromagnet for $H/J=0, 2.9$, and $5.0$ at $T/T_\text{c}=0.5$.
(d) Masked snapshots of the spins of the G-type antiferromagnet for $H/J=0, 2.9$, and $5.0$ at $T/T_\text{c}=0.5$.
(c) and (d) were drawn by VESTA\cite{Momma-2011}.
}
\end{figure*}

\begin{figure*}
\begin{center}
\includegraphics[scale=1.0]{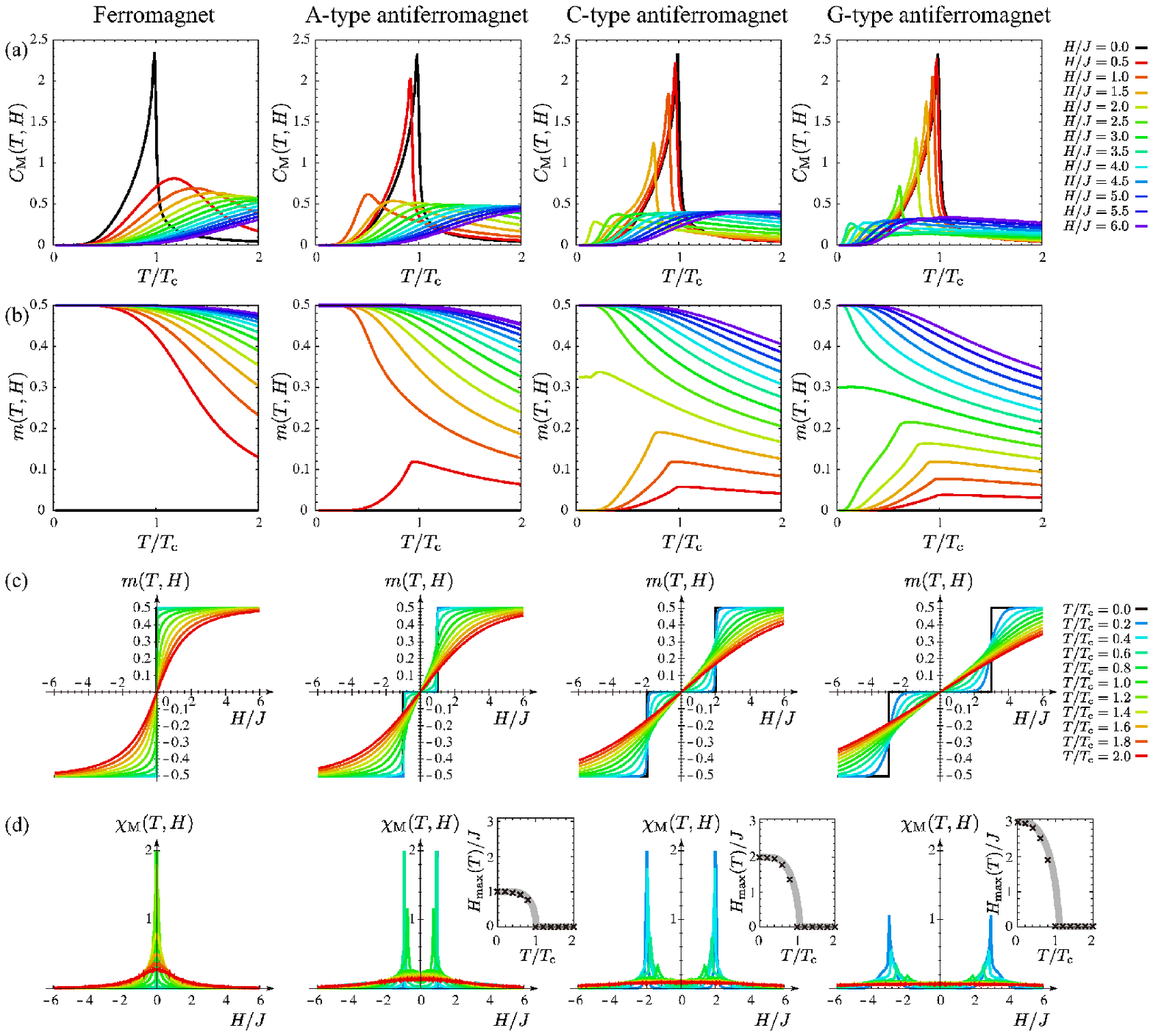} 
\end{center}
\caption{\label{fig:MC_quantities}
(Color online)
(a) Temperature dependence of the magnetic specific heat per spin $C_\text{M}(T,H)$ for $L=16$ under various magnetic fields obtained by the Wang-Landau method.
(b) Temperature dependence of magnetization per spin $m(T,H)$ for $L=16$ under various magnetic fields obtained by the Wang-Landau method.
(c) Magnetic field dependence of $m(T,H)$ at fixed temperatures of $T/T_\text{c}=0.0-2.0$ for $L=16$.
(d) Magnetic field dependence of magnetic susceptibility per spin $\chi_\text{M}(T,H)$ at fixed temperatures of $T/T_\text{c}=0.0-2.0$ for $L=16$.
The Insets show $H_\text{max}(T)/J$ (gray curves) and the peak position of $\chi_{\text{M}}(T,H)$ (black crosses).
}
\end{figure*}

Using the Wang-Landau method, the temperature dependence of the magnetic entropy $S_\text{M}(T,H)$ for $L=16$ is obtained (Fig.~\ref{fig:MC_entropy}).
The magnetic entropies for $L=8$, $12$, and $16$ collapse within the line width in Fig.~\ref{fig:MC_entropy},
thus we use a lattice size of $L=16$ throughout this paper.
In the ferromagnet, the magnetic entropy decreases as the magnetic field increases for any temperature.
The same behavior is observed in the paramagnetic phase above $T_\text{c}$ in antiferromagnets.
In contrast, the magnetic entropy behavior in antiferromagnets below the N\'eel temperature differs from the behavior in the ferromagnet.
The magnetic entropy in antiferromagnets reaches a maximum value at finite $H$ below $T_\text{c}$, whereas in the ferromagnet, the maximum magnetic entropy at a given $T$ is achieved when $H=0$.
Let $H_\text{max} (T)$ be the magnetic field at which the magnetic entropy reaches its maximum value at $T$. 
The insets of Fig.~\ref{fig:MC_entropy} show the temperature dependence of $H_\text{max}(T)$ for antiferromagnets.
As the temperature increases, $H_\text{max}(T)$ monotonically decreases.
In addition, a large residual magnetic entropy of about $\ln 2/2$ is observed in the G-type antiferromagnet with $H/J=3.0$.
The large residual magnetic entropy indicates the existence of macroscopically degenerate ground states, as found in frustrated magnetic systems\cite{Vannimenus-1977,Liebmann-1986,Harris-1997,Kobayashi-1998,Bramwell-2001,Matsuhira-2002,Matsuhira-2002,Udagawa-2002,Yoshioka-2004,Diep-2005,Moessner-2006,Tanaka-2007,Tahara-2007,Andrews-2009,Ogitsu-2010,Tanaka-2010,Tanaka-2012,Chern-2013}.

Next, we consider the microscopic origin of nonzero $H_\text{max} (T)$ in antiferromagnets compared with the ferromagnet. 
Figures~\ref{fig:snapshot} (a) and (b) show the magnetic field dependence of the magnetic entropy at $T/T_\text{c}=0.5$ for the ferromagnet and the G-type antiferromagnet, respectively.
In the ferromagnet, the magnetic entropy decreases as the magnetic field increases because the magnetic field reinforces the ferromagnetic order. 
In contrast, for the G-type antiferromagnet, the magnetic field dependence of the magnetic entropy has a peak at $H_{\text{max}}(T) (\neq 0)$. 
In antiferromagnets, the ordered magnetic structure is destroyed by the magnetic field. 
Thus, the magnetic entropy increases as the magnetic field increases below $H_\text{max}(T)$. 
When we apply a strong magnetic field greater than $H_{\text{max}}(T)$, the spin structure becomes a saturated ferromagnetic structure. 
Thus, the magnetic entropy decreases as the magnetic field increases above $H_{\text{max}}(T)$.
Figure~\ref{fig:snapshot} (c) shows snapshots of the spin configuration of the G-type antiferromagnet at $T/T_{\text{c}}=0.5$.
Because the temperature is lower than the N\'eel temperature, the antiferromagnetic ordered state appears at $H=0$.
To represent the antiferromagnetic ordered structure more clearly, masked snapshots of spin configuration are also shown (Fig.~\ref{fig:snapshot} (d)).
In the masked snapshots, the local gauge transformation $s_i^z \to -s_i^z$ is applied to any spins in one of two sublattices in the G-type antiferromagnet.
Thus, most spins in the masked snapshot at $H/J=0$ are the same color, which indicates that the structure is almost completely antiferromagnetically ordered.
The snapshot of spin configuration at $H/J=2.9$ which is near $H_{\text{max}}(T) (\lesssim 2.9J)$ is shown in the middle panel of Fig.~\ref{fig:snapshot} (c).
The spin structure is almost random, which can be also confirmed in the masked snapshot of the spin configuration (middle panel of Fig.~\ref{fig:snapshot} (d)).
This is a magnetic field induced disordered state.
Near the magnetic field value, the magnetic entropy should be large.
The snapshot of the spin configuration at $H/J=5.0$, which is larger than $H_{\text{max}}(T)$, is shown in the right panel of Fig.~\ref{fig:snapshot} (c).
The spins are almost parallel to the magnetic field at $H/J=5.0$.
As the magnetic field increases from $H_\text{max} (T)$, the saturated ferromagnetic structure appears and the magnetic entropy decreases.


Here, we consider the relation between the behavior of the magnetic entropy and measurable physical quantities.
Figures~\ref{fig:MC_quantities} (a) and (b) show the temperature dependence of the magnetic specific heat $C_\text{M}(T,H)$ and magnetization $m(T,H)$ under various magnetic fields, which are obtained by the Wang-Landau method using Eqs.~(\ref{eq:specificheat}) and (\ref{eq:mag}).
Let us discuss why nonzero $H_{\text{max}}(T)$ exists in antiferromagnets from the behaviors of $C_{\text{M}}(T,H)$ and $m(T,H)$.
The magnetic entropy $S_\text{M} (T,H)$ is calculated from $C_\text{M} (T,H)$ and $m (T,H)$ as
\begin{align}
S_\text{M} (T,H) &= \int_0^{T} \frac{C_\text{M} (T',H)}{T'} dT' + S_\text{M} (0,H),
\label{eq:S-C} \\
S_\text{M} (T,H) &= \int_{0}^H \left( \frac{\partial m (T,H')}{\partial T} \right)_{H'} d H' + S_\text{M} (T,0). \label{eq:S-M}
\end{align}
First, we focus on the magnetic specific heat below $T_{\text{c}}$ (Fig.~\ref{fig:MC_quantities} (a)).
The second term in Eq.~(\ref{eq:S-C}) is the residual magnetic entropy and should always be zero in the ferromagnet.
Since the integrand in the first term in Eq.~(\ref{eq:S-C}) satisfies the inequality
$C_{\text{M}}(T,H_1)/T > C_{\text{M}}(T,H_2)/T$ for any $T(< T_\text{c})$, $H_1$, and $H_2$ with $H_1 < H_2$,
it follows that $S_{\text{M}}(T,H_1)>S_{\text{M}}(T,H_2)$; that is, $H_{\text{max}}(T)$ is always zero below the Curie temperature.
As shown in Fig.~\ref{fig:MC_entropy}, $H_\text{max}(T\to 0)/J=1,2,3$ for the A-, C-, and G-type antiferromagnets, respectively, and $H_\text{max}(T)$ monotonically decreases as the temperature increases. 
Here we consider a magnetic field lower than $H_\text{max}(T \to 0)$. 
In this region, because the residual magnetic entropy is zero as it is for the ferromagnet, it is sufficient to compare the integrand in the first term in Eq.~(\ref{eq:S-C}).
Figure~\ref{fig:MC_quantities} (a) shows that there is a region in which the inequality 
$C_{\text{M}}(T,H_1)/T < C_{\text{M}}(T,H_2)/T$ is satisfied for $H_1<H_2$ below $T_\text{c}$.
In this region, the magnetic entropy increases with the magnetic field.
Thus, in antiferromagnets, $H_{\text{max}}(T)$ is a finite value below the transition temperature.
Next we consider the relation between the magnetic entropy and the magnetization (Fig.~\ref{fig:MC_quantities} (b)).
In the ferromagnet, the magnetization decreases as the temperature increases.
Since the inequality $S_\text{M}(T,H_1) > S_\text{M}(T,H_2)$ is satisfied for any values of $T$, $H_1$, and $H_2$ with $H_1 < H_2$, the value of $H_{\text{max}}(T)$ should be zero for all temperatures from Eq.~(\ref{eq:S-M}).
In contrast, in antiferromagnets below $T_\text{c}$, because $\left( \frac{\partial m (T,H')}{\partial T}\right)_{H'}$ can be positive in the small $H$ region (Fig.~\ref{fig:MC_quantities} (b)), Eq.~(\ref{eq:S-M}) shows that the magnetic entropy can increase with the magnetic field.
Thus, $H_{\text{max}}(T)$ is a finite value below $T_\text{c}$.
The peak position in the temperature dependence of the magnetization under $H$ (Fig.~\ref{fig:MC_quantities} (b)) corresponds to the temperature, such that $H=H_{\text{max}}(T)$.
In other words, the sign of $\left( \frac{\partial m (T,H')}{\partial T}\right)_{H'}$ changes at temperature $T$, such that $H'=H_\text{max}(T)$.

Finally, we show the magnetic field dependence of the magnetization in Fig.~\ref{fig:MC_quantities} (c).
The transition between the antiferromagnetic ordered state, where $m(T,H)=0$, and the saturated ferromagnetic state, where $m(T,H)=0.5$, is observed in antiferromagnets below $T_{\text{c}}$.
The magnetic susceptibility, defined as $\chi_\text{M}(T,H)=\left( \frac{\partial m(T,H)}{\partial H}\right)_T$, is shown in Fig.~\ref{fig:MC_quantities} (d), in which the peak positions correspond to $H_{\text{max}}(T)$ (inset of Fig.~\ref{fig:MC_quantities} (d)).
Note that the magnetic susceptibility peak positions are the metamagnetic transition points.


\section{Magnetic refrigeration efficiency} \label{sec:MCE}

In this section, we consider the magnetic refrigeration efficiency of the ferromagnet and the A-, C-, and G-type antiferromagnets for the isothermal demagnetization process and the adiabatic magnetization process. 
We compare the magnetic refrigeration efficiency of the proposed protocol reported in Ref.~\onlinecite{Tamura-2014} with that of the conventional protocol.

\begin{figure}[b]
\begin{center}
\hspace{10mm} \includegraphics[scale=1.0]{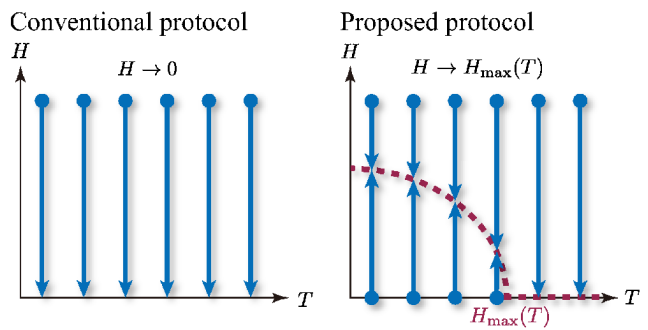} 
\end{center}
\caption{\label{fig:HmaxT}
(Color online)
(Left) Schematic of the conventional protocol, which is suitable for ferromagnets and paramagnets in the isothermal demagnetization process. 
(Right) Schematic of the proposed protocol, which is suitable for antiferromagnets in the isothermal demagnetization process, proposed in Ref.~\onlinecite{Tamura-2014}.
Note that for the adiabatic magnetization process, the direction of the arrows is reversed.
}
\end{figure}

\begin{figure*}
\begin{center}
\includegraphics[scale=1]{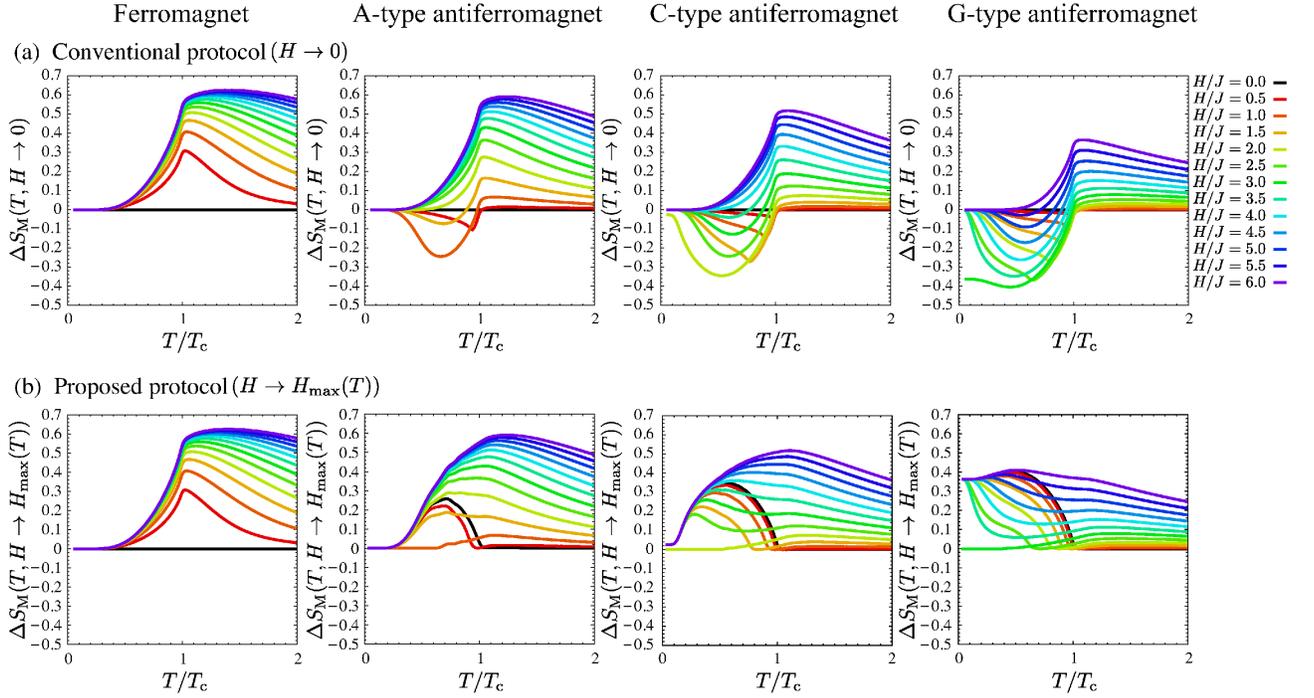} 
\end{center}
\caption{\label{fig:MC_deltaS}
(Color online)
(a) Isothermal magnetic entropy change as a function of temperature under the conventional protocol ($H \to 0$) for $L=16$.
(b) Isothermal magnetic entropy change as a function of temperature under the proposed protocol ($H \to H_\text{max}(T)$) for $L=16$.
}
\end{figure*}

In the isothermal demagnetization process under the conventional protocol, the magnetic field is changed from finite $H$ to zero (left panel of Fig.~\ref{fig:HmaxT}). 
This protocol is efficient when $S_\text{M}(T,H)$ decreases as $H$ increases at a given temperature, which is realized in ferromagnets and paramagnets.
The conventional protocol has been used in most previous studies.
In contrast, Ref.~\onlinecite{Tamura-2014} proposed the protocol that $\Delta S_\text{M}$ given by Eq.~(\ref{eq:def_DSM}) is the maximum value.
Thus, the maximum amount of heat can be absorbed using the proposed protocol shown in the right panel of Fig.~\ref{fig:HmaxT}. 
The magnetic field is changed from $H$ to $H_\text{max} (T)$ (purple dotted curve). 
$H_\text{max} (T)$ is a finite value below $T_\text{c}$, whereas $H_\text{max} (T)=0$ above $T_\text{c}$ in antiferromagnets (insets of Fig.~\ref{fig:MC_entropy}). 
The proposed protocol is more efficient than the conventional protocol when the $H$-dependence of $S_\text{M}(T,H)$ has a peak at finite $H$, as in antiferromagnets (Fig.~\ref{fig:snapshot} (b)). 
Note that because $H_\text{max} (T)=0$ for all temperatures in the ferromagnet, the proposed protocol includes the conventional protocol for ferromagnets and paramagnets.

In Ref.~\onlinecite{Tamura-2014}, the magnetic refrigeration efficiency is only considered with respect to the amount of heat absorption during the isothermal demagnetization process.
In this work, we examine the magnetic refrigeration efficiency in the isothermal demagnetization and adiabatic magnetization processes.
We show that the protocol proposed in Ref.~\onlinecite{Tamura-2014} is useful for obtaining high magnetic refrigeration efficiencies in the two typical processes.

\subsection{Isothermal demagnetization process}
\label{sec:MCE_IDP}

In this subsection, we focus on the isothermal demagnetization process (Fig.~\ref{fig:schematic_S} (a)).
Figures~\ref{fig:MC_deltaS} (a) and (b) are the temperature dependences of $\Delta S_\text{M}(T,H \to 0)$ under the conventional protocol and $\Delta S_\text{M}(T, H \to H_\text{max} (T))$ under the proposed protocol\cite{supplemetal_material1}, respectively.
Here, $H$ is the magnetic field at the start point in both protocols.
In the ferromagnet, because $H_\text{max} (T)$ is always zero, both processes are the same. 
Thus, $\Delta S_\text{M} (T,H \to 0)$ is always positive and the magnetic entropy increases.
In contrast, $\Delta S_\text{M} (T, H \to 0)$ can be negative in antiferromagnets below $T_\text{c}$.
In this case, the magnetic entropy decreases under the conventional protocol, which is called the inverse MCE\cite{Tohei-2003,Krenke-2005,Sandeman-2006,Krenke-2007,Ranke-2009} (Fig.~\ref{fig:MC_deltaS} (a)).
The inverse MCE does not appear in the proposed protocol as $\Delta S_\text{M} (T, H \to H_\text{max} (T))$ is always positive (Fig.~\ref{fig:MC_deltaS} (b)) through the definition of $H_{\text{max}}(T)$.
In addition, $ \Delta S_\text{M} (T, H \to H_\text{max} (T)) \ge \Delta S_\text{M} (T, H \to 0)$ is always satisfied and achieves the maximum value.
Thus, the proposed protocol is useful for obtaining a large isothermal magnetic entropy change in antiferromagnets.

\begin{figure}
\begin{center}
\includegraphics[scale=1.0]{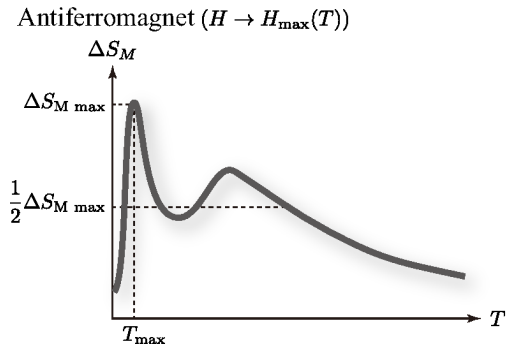} 
\end{center}
\caption{\label{fig:schematic_RCP}
Schematic of the entropy change in the antiferromagnet under the proposed protocol.
}
\end{figure}

Next, we consider the performance of the magnetic refrigeration in the proposed protocol from a different perspective.
Let us consider a refrigerator that transfers heat from a low-temperature reservoir to a high-temperature reservoir when the magnetic field is changed from $H_2$ to $H_1$.
The amount of transferred heat, called the cooling capacity $q$, is given by
\begin{align}
q = \int_{T_{\text{l}}}^{T_{\text{h}}} \Delta S_{\text{M}} (T,H_{2} \to H_{1}) dT,
\label{eq:cc}
\end{align}
where $T_{\text{l}}$ and $T_{\text{h}}$ are the temperatures of the low- and high-temperature reservoirs, respectively.
In ferromagnets, RCP approximately characterizes the cooling capacity when $T_{\text{l}}$ and $T_{\text{h}}$ are set to temperatures such that $\Delta S_{\text{M}}(T_\text{l},H_{2} \to H_{1})=\Delta S_{\text{M}}(T_\text{h},H_{2}\to H_{1})=\frac{1}{2}\Delta S_{\text{M\, max}}$, as shown in Fig.~\ref{fig:schematic_S} (a).
In fact, the RCP value is nearly $4/3$ times the cooling capacity in ferromagnets\cite{Gschneidner-2000}.
However, the RCP is less suitable for measuring the performance of the magnetic refrigeration under the proposed protocol.
In antiferromagnets, the temperature dependence of $\Delta S_{\text{M}}$ under the proposed protocol has, in some cases, more than two temperatures at which $\Delta S_{\text{M}}=\frac{1}{2}\Delta S_{\text{M\, max}}$ (Fig.~\ref{fig:schematic_RCP}).
This situation differs qualitatively from $\Delta S_{\text{M}}$ in ferromagnets or paramagnets shown in Fig.~\ref{fig:schematic_S} (b).
Thus, we introduce a new measure, the TCP, to consider the efficiency of the magnetic refrigeration under the proposed protocol more appropriately. 
The TCP is defined as
\begin{align}
\text{TCP} &= \int_{0}^{\infty} \Delta S_{\text{M}}(T,H_{2}\to H_{1}) \Theta(\Delta S_{\text{M}}(T,H_{2}\to H_{1})) d T, \label{eq:TCP} \\
\Theta(x) &=
\begin{cases}
	0 \quad (x < 0)\\
	1 \quad (x \ge 0)
\end{cases}.
\end{align}
The TCP characterizes the whole potential of the cooling power of the target material.
It is a natural quantity, because the TCP value is almost $3/2$ times the RCP value in the ferromagnet.
A more detailed explanation of the TCP and the calculation method is given in Appendix~\ref{sec:TCP}.

\begin{figure*}
\begin{center}
\includegraphics[scale=1]{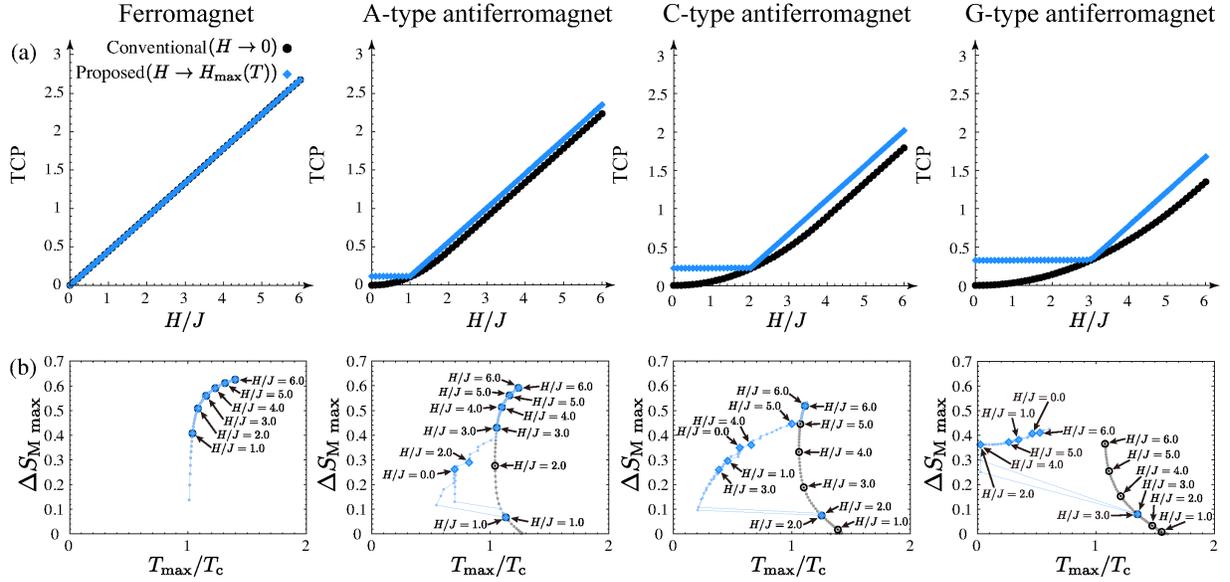} 
\end{center}
\caption{\label{fig:TCP}
(Color online)
(a) Magnetic field at the start point $H$ dependence of TCP under the conventional protocol (black points) and under the proposed protocol (blue points) for $L=16$.
(b) Relation between the maximum magnetic entropy change $\Delta S_\text{M max}$ and the temperature $T_\text{max}$ at which $\Delta S_{\text{M}}= \Delta S_{\text{M\, max}}$ under the conventional protocol (black points) and under the proposed protocol (blue points) for a lattice size of $L=16$.
The lines between points are guide to the eye.
}
\end{figure*}

We consider the TCP of the conventional protocol ($H \to 0$) and that of the proposed protocol ($H \to H_{\text{max}}(T)$).
Figure~\ref{fig:TCP} (a) shows the TCP as a function of $H$ which is the magnetic field at the start point of the given protocols.
In the ferromagnet, the TCPs of both protocols are the same because both protocols are equivalent.
However, for the antiferromagnets, the TCP of the proposed protocol is greater than that of the conventional protocol for any $H$.

Figure~\ref{fig:TCP} (b) shows the relation between the maximum magnetic entropy change $\Delta S_\text{M max}$ and the temperature $T_\text{max}$, at which $\Delta S_{\text{M}}=\Delta S_{\text{M\, max}}$ (Fig.~\ref{fig:schematic_S} (b) and Fig.~\ref{fig:schematic_RCP}) for various $H$.
Using the conventional protocol, $\Delta S_\text{M max}$ monotonically increases with $H$ regardless of the magnetic ordered structure.
Furthermore, $T_\text{max}$ is always larger than $T_\text{c}$, and $T_\text{max}$ monotonically increases in the ferromagnet.
In antiferromagnets, $T_{\text{max}}$ decreases as $H$ increases in the small $H$ region.
However, because $T_{\text{max}}$ should be infinite within the limits of $H \to \infty$, $T_{\text{max}}$ should increase with $H$ in the large $H$ region.
This is not observed using the mean-field analysis, which is discussed in Appendix~\ref{sec:MF}.
In contrast, complicated relation between $\Delta S_\text{M max}$ and $T_\text{max}$ are observed in antiferromagnets using the proposed protocol.
In fact, there is the case that $T_\text{max}$ is smaller than $T_\text{c}$.
We confirm that $\Delta S_\text{M max}$ obtained in the proposed protocol is the same or larger than that in the conventional protocol at a given value of $H$.
In antiferromagnets,
$H_\text{max} (T) \neq 0$ for $T<T_\text{c}$ and $H_\text{max} (T) = 0$ for $T>T_\text{c}$ (insets of Fig.~\ref{fig:MC_entropy}).
Thus, when $T_\text{max} > T_\text{c}$, $\Delta S_\text{M max}$ obtained from the conventional protocol and the proposed protocol are the same.

\subsection{Adiabatic magnetization process}

\begin{figure*}
\begin{center}
\includegraphics[scale=1.0]{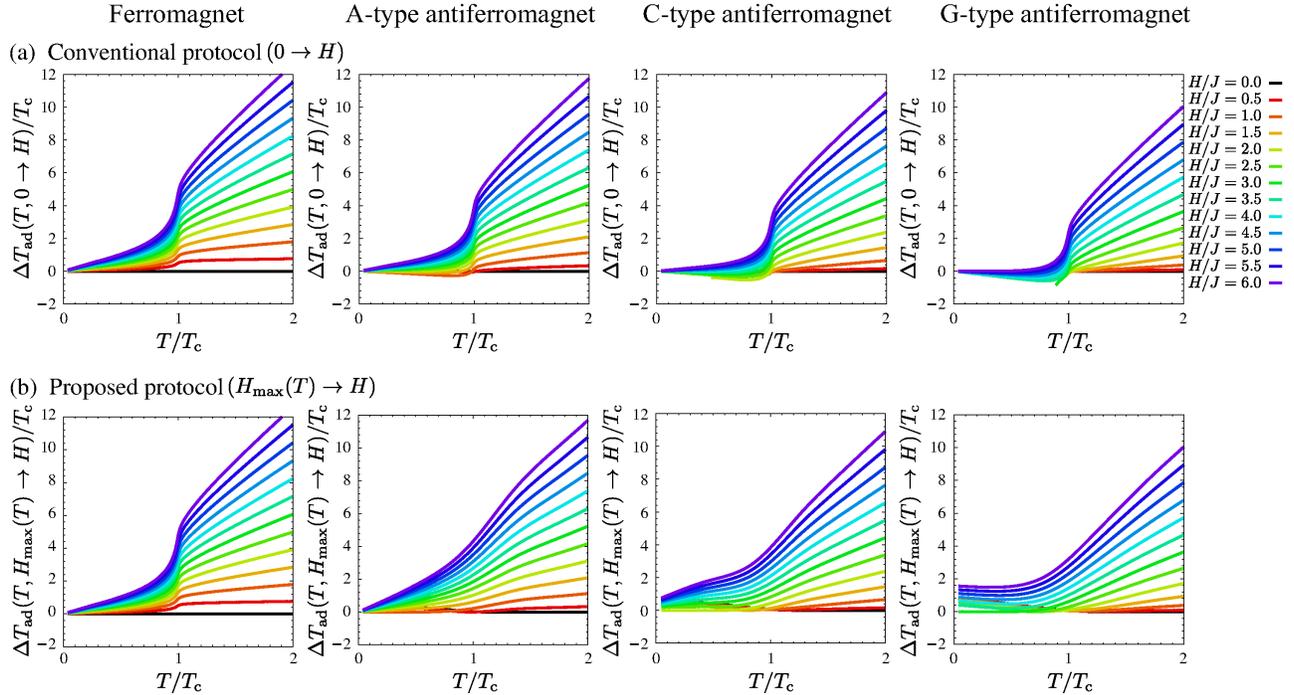} 
\end{center}
\caption{\label{fig:MC_deltaT}
(Color online)
(a) Adiabatic temperature change as a function of temperature at the start point under the conventional protocol ($0 \to H$) for $L=16$.
(b) Adiabatic temperature change as a function of temperature at the start point under the proposed protocol ($H_\text{max}(T) \to H$) for $L=16$.
}
\end{figure*}

In this subsection, we focus on the adiabatic magnetization process.
Let us consider the case that magnetic field is changed from zero to $H$ in the conventional protocol, and from $H_\text{max} (T)$ to $H$ in the proposed protocol (Fig.~\ref{fig:schematic_S} (a)).
Thus, the direction of the arrows in Fig.~\ref{fig:HmaxT} should be reversed.
Figure~\ref{fig:MC_deltaT} (a) is the temperature at the start point dependence of the adiabatic temperature change $\Delta T_\text{ad} (T,0\to H)$ in the conventional protocol.
$\Delta T_\text{ad} (T,0\to H)$ is always positive in the ferromagnet; thus, the temperature of the ferromagnet increases.
However, in antiferromagnets, $\Delta T_\text{ad} (T,0\to H)$ can be negative below $T_\text{c}$.
That is, $H$ is finite, such that $\Delta T_{\text{ad}}(T,0\to H)<0$ for any $T(< T_{\text{c}})$.
In this case, the temperature of antiferromagnets decreases.
For the G-type antiferromagnet with $H/J=3.0$,
$\Delta T_\text{ad} (T,0\to H)$ cannot be well defined below about $T/T_\text{c}=0.9$, because there is no temperature end point in the adiabatic magnetization process. 
The behavior is caused by a large residual magnetic entropy at $H/J=3.0$.
The C-type antiferromagnet with $H/J=2.0$ also shows similar behavior in our calculation. 
Next, we consider the adiabatic temperature change $\Delta T_\text{ad} (T,H_\text{max} (T)\to H)$ under the proposed protocol (Fig.~\ref{fig:MC_deltaT} (b)).
In the proposed protocol, $\Delta T_\text{ad} (T,H_\text{max} (T)\to H)$ is always positive and reaches the maximum value at a given $T$ by the definition of $H_\text{max}(T)$.
Antiferromagnets exhibit a larger adiabatic temperature change in the proposed protocol than in the conventional protocol.
Thus, the proposed protocol is useful for obtaining a large adiabatic temperature change in antiferromagnets. 

\section{Conclusion} \label{sec:conclusion}

We have studied the magnetic refrigeration efficiency in the Ising models of a ferromagnet and A-, C-, and G-type antiferromagnets, which have typical magnetic ordered structures.
The temperature and magnetic field dependences of the magnetic entropy in the Ising models were calculated by the Wang-Landau method, which is a Monte Carlo method.
The obtained magnetic entropy indicates that the protocol proposed in Ref.~\onlinecite{Tamura-2014} achieves the maximum magnetic entropy change in the isothermal demagnetization process and the maximum adiabatic temperature change in the adiabatic magnetization process.
In the proposed protocol, $H_{\text{max}}(T)$ plays an important role, where $H_{\text{max}}(T)$ is the magnetic field at which the magnetic entropy is the maximum value at a given temperature $T$.
$H_{\text{max}}(T)$ is used as the start or end points in the isothermal demagnetization and adiabatic magnetization processes.
The physical meaning of $H_{\text{max}}(T)$ was discussed in terms of measurable physical quantities, such as the magnetic specific heat and the magnetization.
The magnetization process suggests that $H_{\text{max}}(T)$ corresponds to the metamagnetic transition point.

In addition, to estimate the full potential of the cooling power under the proposed protocol, we introduced the new quantity, called total cooling power (TCP).
TCP under the proposed protocol is the same as or larger than that under the conventional protocol in the considered models.
This suggests that the proposed protocol is useful for obtaining the maximum amount of heat transfer.

Finally, we emphasize that the proposed protocol can produce the maximum magnetic refrigeration efficiency in the models we considered and also in other magnetic systems, such as nonferromagnets with an inhomogeneous magnetic ordered structure.
Thus, we believe that this study becomes a fundamental research in the field of magnetic refrigeration.


\section*{Acknowledgment}

We thank Kenjiro Miyano for useful comments and discussions.
R. T., S. T., and H. K. were partially supported by a Grand-in-Aid for Scientific Research (C) (Grant No. 25420698).
In addition, R. T. is partially supported by National Institute for Materials Science.
S. T. is the Yukawa Fellow and his work is supported in part by Yukawa Memorial Foundation.
The computations in the present work were performed on super computers at National Institute for Materials Science, Supercomputer Center, Institute for Solid State Physics, University of Tokyo, and Yukawa Institute of Theoretical Physics. 

\appendix


\section{Cooling power in the ferromagnet} \label{sec:TCP}

\begin{figure}
\begin{center}
\includegraphics[scale=1.0]{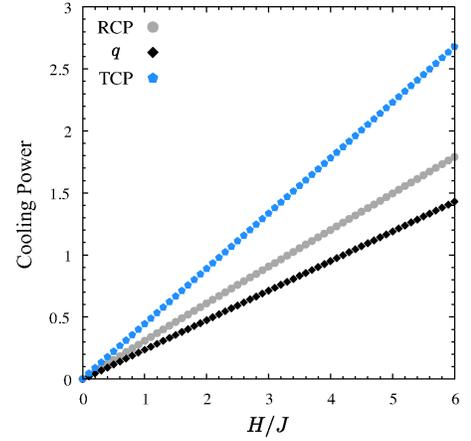} 
\end{center}
\caption{\label{fig:coolingpower}
(Color online) Magnetic field $H$ dependence of RCP (gray circles), cooling capacity $q$ (black diamonds), and TCP (blue pentagons) when the magnetic field is changed from $H$ to zero.
}
\end{figure}

In this appendix, we consider the relation among the relative cooling power (RCP), the cooling capacity $q$, and the total cooling power (TCP) in the ferromagnet.
The definitions of these quantities are given by Eqs.~(\ref{eq:RCP}), (\ref{eq:cc}), and (\ref{eq:TCP}).
The $H$-dependence of these quantities obtained by the Wang-Landau method is shown in Fig.~\ref{fig:coolingpower}, and these quantities almost linearly increases with $H$.
The integration range in the definition of TCP given by Eq.~(\ref{eq:TCP}) is from $0$ to $\infty$.
Practically, it is convenient to divide the integration range into two parts.
We numerically integrate the magnetic entropy change $\Delta S_{\text{M}}$ obtained by the Wang-Landau method from $T=0$ to $20T_{\text{c}}$ and derive $\Delta S_{\text{M}}$ by the single-site approximation from $T=20T_{\text{c}}$ to $\infty$.
At $T\gg T_{\text{c}}$, the single-site approximation is valid because the correlation effect can be ignored.
The Hamiltonian of the single-site approximation is defined as
\begin{align}
\mathcal{H}_\text{single} = - H s^z, \ \ \ \ \ s^z = \pm \frac{1}{2}.
\end{align}
The partition function is given by $Z_\text{single} = \text{Tr} \exp [- \beta \mathcal{H}_\text{single}]$, and the Helmholtz free energy is calculated by $F_\text{single} = - T \ln Z_\text{single}$.
Here, Tr is the summation over all possible states.
From $F_\text{single}$, we can obtain the magnetic entropy
\begin{align}
S_\text{M} (T,H) &= - \frac{\partial F_\text{single}}{\partial T}  \\
&= \ln 2 + \ln \cosh \left( \frac{\beta H}{2} \right) - \frac{\beta H}{2} \tanh \left( \frac{\beta H}{2} \right).
\end{align}
Then, the magnetic entropy change under the single-site approximation is given as 
\begin{align}
\Delta S_\text{M} (T,H \to 0) &= S_\text{M} (T,0) - S_\text{M} (T,H) \\
&=  - \ln \cosh \left( \frac{\beta H}{2} \right) + \frac{\beta H}{2} \tanh \left( \frac{\beta H}{2} \right),
\end{align}
and the contribution in the high temperature region ($T \gg T_{\text{c}}$) is calculated by
\begin{align}
&\int_{T}^{\infty} \Delta S_{\text{M}}(T,H \to 0) \Theta(\Delta S_{\text{M}}(T,H \to 0)) d T \notag \\
& \ \ \ \ \ \ \ \ \ \ \ \ \ \ \ \ \ \ \ \ \ \ \ \ \ \ \ \ \ \ \ \ \ \ \ \ \ =T \ln \cosh \left( \frac{\beta H}{2} \right).
\end{align}
Note that $\Delta S_\text{M} >0$ is always satisfied in the high temperature region ($T \gg T_{\text{c}}$).

Figure~\ref{fig:coolingpower} shows that the TCP value is nearly $3/2$ times the RCP value.
Furthermore, because the relation RCP $\simeq 4q/3$ is satisfied, as explained in Sec.~\ref{sec:MCE_IDP}, we obtain the relation TCP $\simeq 2 q$.
Thus, TCP is a natural quantity for estimating the potential of magnetic refrigeration efficiency.


\section{Mean-field calculation results} \label{sec:MF}

In this appendix, we examine the MCE in the models given by Eq.~(\ref{eq:model}) using a mean-field approximation.
To calculate $S_\text{M}(T,H)$ of the ferromagnet and the A-, C-, and G-type antiferromagnets in a unified way, we consider a system which consists of two sublattices, labeled by $A$ and $B$.
Here, we define the Hamiltonian of the mean-field model as
\begin{align}
\mathcal{H}_\text{MF} = \eta_A \sum_{i \in \text{sublattice} \ A} s_i^z + \eta_B \sum_{i \in \text{sublattice} \ B} s_i^z,
\end{align}
where the first and second sums are over all spins in sublattices $A$ and $B$, respectively,
and $\eta_A$ and $\eta_B$ are variational parameters.
The partition function is given by $Z_\text{MF} = \text{Tr} \exp [- \beta \mathcal{H}_\text{MF}]$,
and the canonical ensemble average of physical quantity $\mathcal{Q}$ can be calculated using $\langle \mathcal{Q} \rangle_\text{MF} = Z_\text{MF}^{-1} \text{Tr} \mathcal{Q} \exp [- \beta \mathcal{H}_\text{MF}]$.
The Helmholtz free energy is given by $F_\text{MF} = - T \ln Z_\text{MF}$.
Let $J_{\mu \nu}$ be the interactions between spins in sublattices $\mu$ and $\nu$ ($\mu, \nu = A, B$) and $m_\mu$ be the sublattice magnetization in $\mu$ ($\mu = A, B$).
From the variational principle for the Helmholtz free energy, the Bogoliubov inequality is given as
\begin{align}
F \le &F_\text{MF} + \langle \mathcal{H} - \mathcal{H}_\text{MF} \rangle_\text{MF} \label{eq:Bogoliubov} \\
= &-\frac{NT}{2} \left[ \log \left( 2 \cosh \frac{1}{2} \beta \eta_A \right) + \log \left( 2 \cosh \frac{1}{2} \beta \eta_B \right) \right] \notag \\
&- \frac{N}{4} \left[ z_{AA} J_{AA} m_A^2 +  2 z_{AB} J_{AB} m_A m_B + z_{BB} J_{BB} m_B^2 \right] \notag \\
&- \frac{N}{2} (H-\eta_A) m_A - \frac{N}{2} (H-\eta_B) m_B, \label{eq:variation} \\
&m_A = \frac{1}{2} \tanh \frac{1}{2} \beta \eta_A, \ \ \ \ m_B = \frac{1}{2} \tanh \frac{1}{2} \beta \eta_B,
\end{align}
where $z_{\mu \nu}$ $(\mu, \nu = A, B)$ are the number of nearest-neighbor spins in sublattice $\nu$ surrounding a spin in sublattice $\mu$.
In this study, $z_{\mu \nu}=z_{\nu \mu}$ is satisfied in all cases considered, although $z_{\mu \nu} \neq z_{\nu \mu}$ in general magnetic systems\cite{Ohkoshi-1999}.
The parameters for each magnetic ordered structure are summarized in Table~\ref{tab:nn_parameter}.
The equilibrium state is given such that the equality is satisfied in Eq.~(\ref{eq:Bogoliubov}).

\begin{table}
\caption{\label{tab:nn_parameter}
Interactions $J_{\mu \nu}$ and the number of the nearest neighbor spins in sublattice $\nu$ surrounding a spin in sublattice $\mu$, $z_{\mu \nu}$ ($\mu, \nu = A, B$), for each magnetic ordered structure.
}
\begin{tabular}{l|ccc|ccc}
\hline\hline
 &  \ $J_{AA}$  &  $J_{AB}$  &  $J_{BB}$ \  & \  $z_{AA}$  &  $z_{AB}$  &  $z_{BB}$  \\
\hline
Ferromagnet  & $0$ & $+J$ & $0$ & $0$ & $6$ & $0$ \\
A-type antiferromagnet  & $+J$ & $-J$ & $+J$ & $4$ & $2$ & $4$ \\
C-type antiferromagnet  & $+J$ & $-J$ & $+J$ & $2$ & $4$ & $2$ \\
G-type antiferromagnet  & $0$ & $-J$ & $0$ & $0$ & $6$ & $0$ \\
\hline\hline
\end{tabular}
\end{table}

\begin{figure*}
\begin{center}
\includegraphics[scale=1.0]{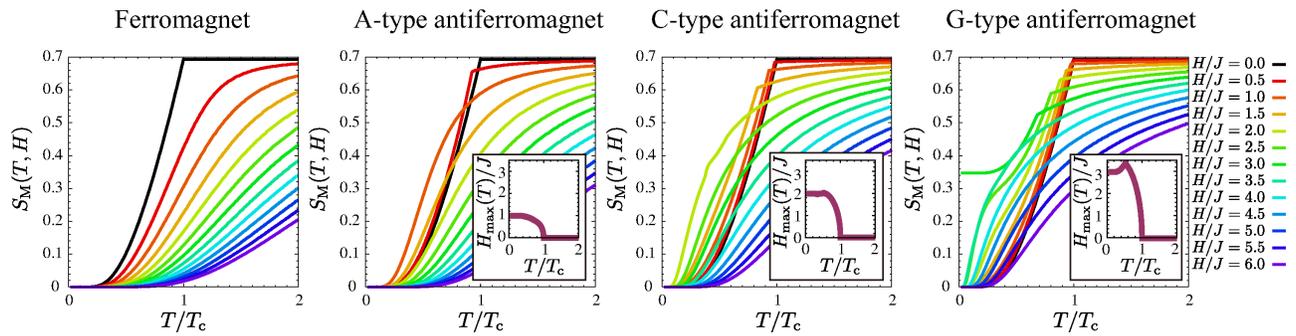} 
\end{center}
\caption{\label{fig:MF}
(Color online)
Temperature dependence of magnetic entropy per spin $S_\text{M}(T,H)$ obtained from mean-field analysis.
The insets show the temperature dependence of $H_\text{max} (T)$ at which the magnetic entropy reaches the maximum value at a given $T$.
}
\end{figure*}

The magnetic entropy per spin is calculated by
\begin{align}
S_\text{M} (T,H) &= - \frac{1}{N} \frac{\partial }{\partial T} \left[ F_\text{MF} + \langle \mathcal{H} - \mathcal{H}_\text{MF} \rangle_\text{MF} \right] \\
&= \frac{1}{2} \left[ \log \left( 2 \cosh \frac{1}{2} \beta \eta_A \right) + \log \left( 2 \cosh \frac{1}{2} \beta \eta_B \right) \right] \notag \\
&\ \ \ \ - \frac{1}{2T} ( \eta_A m_A + \eta_B m_B) \label{eq:ent_MF}.
\end{align}
Figure~\ref{fig:MF} shows the temperature dependence of the magnetic entropy for various magnetic fields using Eq.~(\ref{eq:ent_MF}).
This is similar to the magnetic entropy behavior obtained by the Wang-Landau method (Fig.~\ref{fig:MC_entropy}), except that $S_\text{M}(T,H=0)=\ln 2$ is above the transition temperature $T_{\text{c}}$.
In this naive mean-field analysis, the specific heat at $H=0$ is zero above the transition temperature because the removal of the thermal fluctuation effect is excessive.
Therefore, $T_{\text{c}}$ calculated by the mean-field approximation is greater than the true transition temperature.
The transition temperature of the mean-field model is $T_\text{c}/J=1.5$, whereas the true transition temperature is $T_\text{c}/J=1.127 \cdots$, as explained in Sec.~\ref{sec:model}.
The insets of Fig.~\ref{fig:MF} show $H_\text{max}(T)$ obtained by the mean-field analysis.
$H_\text{max}(T)$ shows nonmonotonic behavior as the temperature varies in the low temperature region, whereas the Wang-Landau method shows that $H_\text{max}(T)$ monotonically decreases as $T$ increases (insets of Fig.~\ref{fig:MC_entropy}).
Similar behavior is observed at the phase boundary of the antiferromagnetic Ising model.
The phase boundary in the low temperature region using the naive mean-field analysis qualitatively differs from that obtained by the Monte Carlo method\cite{Garrett-1951,Shirley-1977,Landau-1977}.
However, as the precision of approximation increases, (\textit{i.e.}, higher order fluctuation is included), the approximated phase boundary approaches the true phase boundary\cite{Ziman-1951,Bienenstock-1966,Wentworth-1993}.
Therefore, the nonmonotonicity of $H_\text{max}(T)$ is an artifact that arises from the naive mean-field analysis.
Using the data shown in Fig.~\ref{fig:MF}, $\Delta S_\text{M}$ in the isothermal demagnetization process and $\Delta T_\text{ad}$ in the adiabatic magnetization process for the conventional and proposed protocols can be obtained (Supplemental Material~\cite{supplemetal_material2}).

\end{document}